\documentclass{INTERSPEECH2023}
\usepackage[T2C,T2A]{fontenc}
\usepackage[utf8]{inputenc}
\usepackage[russian,english]{babel}
\usepackage{substitutefont}
\substitutefont{T2A}{\rmdefault}{Tempora-TLF}

\usepackage{url}

\DeclareTextSymbolDefault{\cyrghk}{T2C}
\DeclareTextSymbolDefault{\cyrnhk}{T2C}
\usepackage{tipa}

\usepackage{hyperref}
\usepackage{multirow,url,booktabs}

\newcommand{\scell}[2][c]{%
  \begin{tabular}[#1]{@{}c@{}}#2\end{tabular}}
\interspeechcameraready

\title{Multilingual Text-to-Speech Synthesis for\\Turkic Languages Using Transliteration}

\name{Rustem Yeshpanov, Saida Mussakhojayeva, Yerbolat Khassanov}
\address{Institute of Smart Systems and Artificial Intelligence, Nazarbayev University, Kazakhstan}
\email{\{rustem.yeshpanov, saida.mussakhojayeva, yerbolat.khassanov\}@nu.edu.kz}

\begin{document}

\maketitle
 
\begin{abstract}
This work aims to build a multilingual text-to-speech (TTS) synthesis system for ten lower-resourced Turkic languages: Azerbaijani, Bashkir, Kazakh, Kyrgyz, Sakha, Tatar, Turkish, Turkmen, Uyghur, and Uzbek.
We specifically target the zero-shot learning scenario, where a TTS model trained using the data of one language is applied to synthesise speech for other, unseen languages. 
An end-to-end TTS system based on the Tacotron 2 architecture was trained using only the available data of the Kazakh language.
To generate speech for the other Turkic languages, we first mapped the letters of the Turkic alphabets onto the symbols of the International Phonetic Alphabet (IPA), which were then converted to the Kazakh alphabet letters.
To demonstrate the feasibility of the proposed approach, we evaluated the multilingual Turkic TTS model subjectively and obtained promising results.
To enable replication of the experiments, we make our code and dataset publicly available in our GitHub repository\footnote{\url{https://github.com/IS2AI/TurkicTTS}\label{ft:github}}. 
\end{abstract}
% \noindent\textbf{Index Terms}: speech synthesis, TTS, Turkic, IPA, lower-resourced, zero-shot, cross-lingual, transfer learning

\section{Introduction}
\label{sec:intro}
%What is the TTS? Why is it important? What is the one of the main problems in TTS?
% Text-to-speech (TTS) synthesis has shown great success on resource-rich languages, where a large amount of labeled data is available [cite].
Text-to-speech (TTS) synthesis has proved successful for higher-resourced languages, for which a large amount of labelled data is available~\cite{DBLP:conf/interspeech/WangSSWWJYXCBLA17,DBLP:conf/icassp/ShenPWSJYCZWRSA18,DBLP:conf/icassp/YamamotoSK20}.
% \footnote{The terms \textit{lower-} and \textit{higher-resourced languages} are used throughout the paper to emphasise the continuum existing across languages in terms of resources available for language technology development.}
% The generated speech are of high quality and sound more human-like, finding its application in many real-world tasks.
%The generated speech is of high quality and sounds human-like, finding its application in many real-world tasks.
The generated speech is of high quality and sounds human-like, finding its use in many commercial applications.
%Importantly, TTS has wide application potential and a substantial social impact, including digital assistants, improved accessibility for people with reading disabilities, speech and vision impairments, to name a few.
Importantly, TTS has substantial social impact on assistive technologies for people with disabilities, such as speech and vision impairments, making it an essential speech processing technology for any language.
% However, for the majority of world languages TTS systems remain inaccessible.
However, for the majority of the world's languages, TTS systems remain inaccessible.
% This is mostly due to the lack of labelled data of sufficient size since the TTS data collection is considered highly laborious [cite].
This is mainly due to the lack of labelled data of sufficient size, as TTS data collection is considered highly laborious~\cite{DBLP:conf/interspeech/ChenTYL19}.

% \begin{figure}[t]
%     \centering
%     \includegraphics[width=0.99\linewidth, trim={0cm 6.8cm 0cm 0cm}, clip=true]{Figures/Map-TurkicLanguagesGroups.png}
%     \caption{Four branches of Turkic languages: red - Oghuz, green - Karluk, cyan \& blue - Kipchak, dark blue - Siberian.}
% \label{fig:map}
% \end{figure}

%How does this paper is going to address this problem?
% In this work, we aim to build a single TTS model supporting ten Turkic languages including Azeri, Bashkir, Kazakh, Kyrgyz, Tatar, Turkish, Turkmen, Uyghur, Uzbek, and Yakut (or Sakha).

In this work, we aim to build a multilingual TTS model supporting ten Turkic languages, including Azerbaijani, Bashkir, Kazakh, Kyrgyz, Sakha, Tatar, Turkish, Turkmen, Uyghur, and Uzbek.
% In this work, we aim to build a multilingual TTS model supporting ten Turkic\footnote{In this work, the term Turkic will be used to collectively refer to addressed ten languages.} languages, including Azerbaijani, Bashkir, Kazakh, Kyrgyz, Sakha, Tatar, Turkish, Turkmen, Uyghur, and Uzbek.
% These languages are considered lower-resourced, and to the best of our knowledge, this work will be the first end-to-end (E2E) TTS system building attempt for most of them.
These languages are considered lower-resourced, and, to the best of our knowledge, this work is the first attempt to develop an end-to-end (E2E) TTS system for most of them.
% At large, our work is motivated by the recent work of Mussakhojayeva et al.~\cite{mussakhojayeva21_interspeech} who addressed the data scarcity in Kazakh language by developing a large-scale and open-source speech corpus named KazakhTTS2.
Our study became feasible thanks to the recent work of Mussakhojayeva et al.~\cite{mussakhojayeva2022kazakhtts2}, who have addressed data scarcity in the Kazakh language, by developing a large-scale and open-source speech corpus called KazakhTTS2.
% The KazakhTTS2 consists of 5 speakers covering over 270 hours of high-quality transcribed speech data.
%KazakhTTS2 consists of over 270 hours of high-quality transcribed speech data narrated by five speakers.
%Recently, Mussakhojayeva et al.~\cite{mussakhojayeva21_interspeech} presented KazakhTTS2, a large-scale and open-source Kazakh speech corpus for TTS applications.
%The KazakhTTS2 dataset consists of 5 voices covering over 270 hours of high-quality transcribed speech data.
% This corpus unlocked new opportunities for previously understudied family of Turkic languages.
This corpus unlocks new opportunities for the hitherto understudied—in regard to spoken language technology—family of Turkic languages.
% Thus, by taking the advantage of KazakhTT2, we study the knowledge transfer from Kazakh to other Turkic langauges.
% Thus, by taking advantage of KazakhTTS2, we investigate knowledge transfer from Kazakh to other Turkic languages.
Thus, by taking advantage of KazakhTTS2, we investigate the Kazakh transliteration of other Turkic languages.
%In this work, by taking the advantage of KazakhTT2, we aim to build a single TTS model supporting ten Turkic languages including Azeri, Bashkir, Kazakh, Kyrgyz, Tatar, Turkish, Turkmen, Uyghur, Uzbek, and Yakut (or Sakha).
%Specifically, we address ten Turkic languages, where we aim to build a single Turkic TTS model using KazakhTTS2 dataset and cross-lingual transfer learning approaches. 
Specifically, to bring together the target languages under the same input space, we employed the International Phonetic Alphabet (IPA)~\cite{international1999handbook}, manually mapping all letters of the Turkic alphabets onto their IPA representations.

% The addressed ten Turkic languages form a comprehensive language family with over 150 million L1 speakers.
The ten Turkic languages under consideration form a comprehensive language family with over 150 million native speakers~\cite{wiki}.
% L1 speakers: Azeri - 9.2M, Bashkir - 1.2 M, Kazakh - 12.6M, Kyrgyz - 5.1M, Tatar - 5.3M, Turkish - 82.2M, Turkmen - 6.6M, Uyghur - 10.4M, Uzbek - 32.4M, Yakut - 0.5M
% They are spoken across a wide geographical area stretching from the Balkans through Central Asia to northeast Siberia.
% These languages can be classified into four branches\footnote{We did not cover any language from the remaining two branches (Chuvash and Khalaj).} as depicted in Figure~\ref{fig:map}.
% Spoken across a wide geographical area stretching from the Balkans through Central Asia to northeastern Siberia, these languages can be divided into four branches\footnote{We did not include languages from the remaining two branches (Chuvash and Khalaj).}, as depicted in Figure~\ref{fig:map}.
Spoken across a wide geographical area stretching from the Balkans through Central Asia to northeastern Siberia, these languages can be divided into four branches, as shown in Table~\ref{tab:characteristics}.
% Additionally, these languages share a wide range of common linguistic features such as vowel harmony, agglutination, subject-object-verb order, and lack of grammatical gender, where the intensity of each feature might vary from language to language.
The languages share a wide range of common linguistic features, such as vowel harmony, extensive agglutination, subject-object-verb order, and the absence of grammatical gender and articles, although the intensity of each feature may vary from language to language.
% We believe that the linguistic similarities present among Turkic languages should further amplify the efficiency of cross-lingual transfer learning methodologies in speech and text processing tasks [cite].
% We believe that the geographical proximity and similarities in phonology, morphology, and syntax among Turkic languages should further enhance the efficiency of cross-lingual transfer learning methodologies, as was observed in~\cite{DBLP:conf/acl/LinCLLZXRHZMALN19,lauscher2020zero}.
We believe that the geographical proximity and similarities in phonology, morphology, and syntax among Turkic languages should further enhance the efficiency of our approach, as was observed in~\cite{DBLP:conf/acl/LinCLLZXRHZMALN19,lauscher2020zero}.
%The linguistic similarities present among Turkic languages should ease the application of cross-lingual transfer learning methodologies for speech and text processing tasks.
% The detailed characteristics of covered Turkic languages are provided in Table~\ref{tab:characteristics}.
\begin{table}[t]
    \begin{center}
        %\small
        %\renewcommand\arraystretch{1.1}
        \setlength{\tabcolsep}{3.0mm}
        \caption{The characteristics of the ten Turkic languages}\label{tab:characteristics}
        \vspace{-0.2cm}
        \begin{tabular}{llrl}
            \toprule
            \textbf{Language}   & \textbf{Branch}   & \textbf{\scell{Native\\speakers}} & \textbf{\scell{Main writing\\system}}\\
            \midrule
            Azerbaijani         & Oghuz             & 33M                               & Latin\\ 
            Bashkir             & Kipchak           & 1.5M                              & Cyrillic\\ 
            Kazakh              & Kipchak           & 14M                               & Cyrillic\\ 
            Kyrgyz              & Kipchak           & 5M                                & Cyrillic\\
            Sakha               & Siberian          & 0.4M                              & Cyrillic\\ 
            Tatar               & Kipchak           & 5.5M                              & Cyrillic\\ 
            Turkish             & Oghuz             & 83M                               & Latin\\ 
            Turkmen             & Oghuz             & 7M                                & Latin\\ 
            Uyghur              & Karluk            & 11M                               & Perso-Arabic\\ 
            Uzbek               & Karluk            & 27M                               & Latin\\
            \bottomrule
        \end{tabular}
        \vspace{-1.0cm}
        %{\raggedright \textit{Note.} The news source data of speakers F1 and M1 are from KazakhTTS.\par}
    \end{center}
\end{table}

We evaluated the developed multilingual Turkic TTS system based on the Tacotron 2 architecture~\cite{DBLP:conf/icassp/ShenPWSJYCZWRSA18} using subjective tests and obtained promising results for all the languages. 
%The obtained TTS model was evaluated using the subjective mean opinion score (MOS), with promising results for all the languages obtained.
Specifically, we evaluated (1) the overall quality, using the mean opinion score (MOS) measure, (2) the comprehensibility, as well as (3) the intelligibility of the synthesized speech.
%Specifically, for the \textbf{bla1, bla2, and bla3} languages, the MOS value is above \textbf{4.0}, while, for \textbf{bla4, bla5, and bla6}, the MOS value is within \textbf{X-Y}.
% The obtained MOS scores indicate that our TTS model is suitable for most real-world applications.
The obtained results indicate that our TTS model is suitable for most real-world applications.
% For the remaining languages, obtained scores are unsatisfactory, and we discuss the potential reasons in the section below and leave it as a future work.

%For the remaining target languages, the obtained values are unsatisfactory. The possible reasons for and practical solutions to this are elaborated on in the Challenges section and are left for future work.
% \vspace{0.5cm}
% The main contributions of the paper:
The main contributions of the work:
\begin{itemize}
    % \item We propose a zero-shot cross-lingual transfer learning based approach for building multilingual TTS system for Turkic languages. Different from other approaches, our method does not require any data for target languages. To the best of our knowledge, this is the first attempt to build TTS system for some of the addressed languages.
    % \item We propose an approach based on zero-shot cross-lingual transfer learning for building a multilingual TTS system for Turkic languages. Unlike other approaches, our method does not require data for the target languages. To the best of our knowledge, this is the first attempt to build a TTS system for some of the languages addressed (which langs?).
    % \item We investigate an IPA-based approach for building a multilingual TTS system for ten Turkic languages under the zero-shot learning scenario. To the best of our knowledge, for majority of the target languages this is the first attempt to build the E2E TTS system.
    \item We investigate an IPA-based approach to build a multilingual E2E TTS system for ten Turkic languages under the zero-shot learning scenario; %To the best of our knowledge, this is the first attempt to develop an E2E TTS system for most of the target languages.
    % \item We evaluated the utility of a developed IPA-based TTS system using both subjective and objective metrics. The results indicate that obtained system can be used as a strong baseline for future work.
    % \item We have evaluated the utility of the developed Turkic TTS system using both subjective and objective metrics. The results indicate that the developed system can be employed in real-world tasks and used as a strong baseline for future work.
    \item We evaluate the utility of the developed Turkic TTS system by subjectively assessing the overall quality, comprehensibility, and intelligibility of the synthesised speech; %The results indicate that the developed system can be employed in real-world tasks and used as a strong baseline for future work.
    % \item We modified existing IPA to construct Turkic-to-IPA and IPA-to-Turkic converters. These converters can benefit other applications for Turkic languages such as speech recognition and translation.
    % \item The implemented codes, including TTS training and IPA converters, and used dataset are made publicly available in our GitHub repository.
    % \item The implemented codes, including the TTS model and IPA converters, as well as the dataset used, are publicly available in our GitHub repository\textsuperscript{\ref{ft:github}}.
    % \item The implemented codes, including the TTS model and the IPA converters, as well as the dataset used, are made publicly available in our GitHub repository\textsuperscript{\ref{ft:github}}.
    \item The implemented codes, including the TTS model and the IPA converters, as well as the dataset used, are made publicly available in our GitHub repository\textsuperscript{\ref{ft:github}}.
\end{itemize}

% The rest of the paper is structured as follows.
% The rest of the paper is structured as follows:
% Section~\ref{sec:related} briefly reviews the low-resource TTS approaches and prior work on Turkic TTS.
The rest of the paper is structured as follows:
Section~\ref{sec:related} briefly reviews lower-resourced TTS approaches and previous work on Turkic TTS.
% Section~\ref{sec:ipa} explains constructed IPA converters for Turkic languages.
% Section~\ref{sec:methodology} explains the proposed approach, including the IPA converters for the Turkic languages and cross-lingual transfer learning scheme.
Section~\ref{sec:methodology} provides a thorough overview of the proposed pipeline.
% Section~\ref{sec:exp_setup} describes experimental setup, and
Section~\ref{sec:exp_setup} describes the experimental setup, and Section~\ref{sec:results} presents the evaluation results.
% Section~\ref{sec:results} presents evaluation results.
%Section~\ref{sec:results} presents the evaluation results.
% Section~\ref{sec:discuss} discussed important findings and faced challenges.
Section~\ref{sec:discuss} discusses the challenges faced, important findings, and potential future work.
% Section~\ref{sec:conc} concludes this paper and list potential future work.
Section~\ref{sec:conc} concludes this paper.
\vspace{-0.2cm}

\section{Prior work on Turkic TTS}
\label{sec:related}
Turkic languages are generally considered to be lower-resourced, as publicly available linguistic data are limited.
% To address this, recently, several works developed high-quality open-source datasets.
To address this, several recent works have developed high-quality open-source datasets.
% Especially, the trend is observable in Kazakh language with the freely available datasets constructed for the speech and text processing tasks, such as named entity recognition~\cite{yeshpanov2021kaznerd}, speech recognition~\cite{khassanov2020crowdsourced}, and speech synthesis~\cite{mussakhojayeva2022kazakhtts2,mussakhojayeva21_interspeech}.
% This trend can especially be observed for Kazakh with freely available datasets constructed for speech and text processing tasks, such as named entity recognition~\cite{yeshpanov2021kaznerd}, speech recognition~\cite{khassanov2020crowdsourced,DBLP:conf/specom/MussakhojayevaK21,mussakhojayeva2022ksc2}, and speech synthesis~\cite{mussakhojayeva2022kazakhtts2,mussakhojayeva21_interspeech}.
This trend can especially be observed for Kazakh, where freely available datasets have been constructed for speech and text processing tasks, such as named entity recognition~\cite{yeshpanov2021kaznerd}, speech recognition~\cite{khassanov2020crowdsourced,DBLP:conf/specom/MussakhojayevaK21,mussakhojayeva2022ksc2}, and speech synthesis~\cite{mussakhojayeva2022kazakhtts2,mussakhojayeva21_interspeech}.
% In particular, it is important to mention KazakhTTS2 corpus, which enabled our work.
% In particular, it is important to mention the KazakhTTS2 corpus, which made our work possible.
Of particular note is the KazakhTTS2 corpus, which made our work possible.
KazakhTTS2 consists of five voices (three female and two male), with over 270 hours of high-quality transcribed data. 
% The corpus is publicly available, which permits both academic and commercial use.
The corpus is in the public domain and can be used both academically and commercially.

% Another highly studied Turkic language is Turkish with many published works present in literature~\cite{DBLP:conf/interspeech/SalorPD03,gormez2008ttts,turkishTTS1}.
Another very well-studied Turkic language is Turkish, on which there are many published works in the literature~\cite{salor2003implementation,turkishTTS1}.
% For example, one of the earliest works~\cite{DBLP:conf/interspeech/SalorPD03} developed diphone inventory for Turkish to construct diphone based concatenative TTS systems.
One of the earliest works~\cite{salor2003implementation}, for example, developed a diphone inventory for Turkish to construct diphone-based concatenative TTS systems.
% In more recent work, Ergün and Yıldırım~\cite{turkishTTS1} studied whether English data can be used to train Turkish TTS system, where satisfactory results are obtained.
In a more recent work, Ergün and Yıldırım~\cite{turkishTTS1} investigated whether English data could be used as a source language to train a Turkish TTS system and obtained satisfactory results.
% Interestingly, we couldn't find any large-scale and open-source speech corpus dedicated to Turkish speech synthesis\footnote{Should we mention here that we search only papers in English as such a paper might exist in corresponding Turkic language?}.
Despite the abundance of works focusing on Turkish TTS, we were not able to find a large-scale and open-source speech corpus developed for Turkish speech synthesis\footnote{It should be stressed that only English language papers were consulted, for which reason a number of publications might have been left out of the study.}.

% Several works have addressed Uyghur TTS~\cite{silamu2009speech,8659652}, however, these studies are usually conducted using proprietary in-house data.
Several papers have addressed Uyghur TTS~\cite{silamu2009speech,8659652}.
These studies, however, were usually conducted using proprietary, in-house data.
% Moreover, most of the used datasets consists of few hundred utterances, which might be insufficient for building reliable TTS systems.
Moreover, most of the datasets used consist of a few hundred utterances, which might be insufficient for building reliable TTS systems.
% Similarly, we found few works investigating TTS systems for Azerbaijani~\cite{rustamov2014approach}, Tatar~\cite{khusainov2015design}, Uzbek~\cite{akmuradov2021developing}, and Yakut~\cite{leontiev2020improving}.
Similarly, we found only a few papers investigating TTS systems for Azerbaijani~\cite{rustamov2014approach}, Sakha~\cite{leontiev2020improving}, Tatar~\cite{khusainov2015design}, and Uzbek~\cite{akmuradov2021developing}.
% These works mostly focused on conventional TTS approaches, such as concatenative and unit selection.
% These works mostly focused on conventional TTS approaches, such as unit selection and concatenative.
These works mostly focused on conventional TTS approaches, such as unit selection and concatenation.
% Furthermore, the datasets used in these works are either unavailable or unsuitable for building state-of-the-art E2E TTS systems. 
Furthermore, the datasets used in these works are either unavailable or unsuitable for developing state-of-the-art TTS systems. 
% We couldn't find any reasonable work dedicated to Bashkir, Kyrgyz, and Turkmen speech synthesis as existing work focus on specific aspects of TTS rather than complete speech synthesis process.
% We could not find any reasonable work dealing with speech synthesis for Bashkir, Kyrgyz, or Turkmen, as the existing works focus on specific aspects of TTS rather than on the complete speech synthesis process.
We could not find any reasonable work dealing with speech synthesis for Bashkir, Kyrgyz, or Turkmen, as the existing research focuses on specific aspects of TTS rather than on the complete speech synthesis process.
\vspace{-0.2cm}

\section{Methodology} 
\label{sec:methodology}
\begin{figure}[t]
    \centering
    \includegraphics[width=0.9\linewidth, trim={0cm 14.4cm 5cm 0cm}, clip=true]{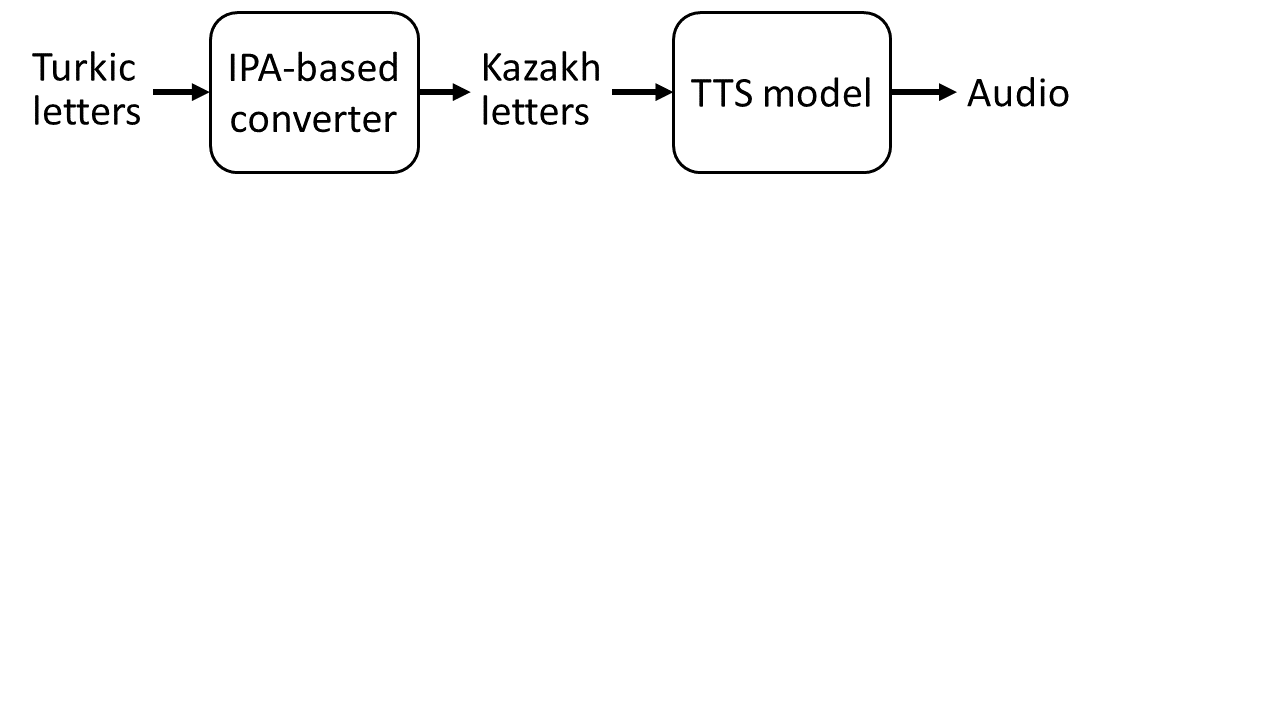}
    \vspace{-0.15cm}
    \caption{The multilingual Turkic TTS system overview}
\label{fig:arch}
\vspace{-0.7cm}
\end{figure}

% The overall architecture of proposed approach is depicted in Figure~\ref{fig:arch}, which consists of two main modules: IPA-based converter and TTS model.
The overall architecture of the proposed approach is shown in Figure~\ref{fig:arch}, which consists of two main modules: an IPA-based converter and a TTS model.
%The system development procedure consists of two main stages: training and generation.
% First, we train a TTS model using Kazakh as a source language, with the letters of the Kazakh alphabet being used as an input sequence.
First, we train a TTS model using Kazakh as a source language, with the letters of the Kazakh alphabet as an input sequence.
%Note that the inputs to the system are letters of Kazakh alphabet.
% The training process is same as in~\cite{mussakhojayeva2022kazakhtts2}, where a mean opinion score (MOS) above four (out of five) is achieved for all voices present in KazakhTTS2.
The training process is the same as in~\cite{mussakhojayeva2022kazakhtts2}, where a MOS of above four (out of five) was achieved for all voices present in KazakhTTS2.
% Specifically, for each voice, we train a separate TTS model (in total five models).
More precisely, for each voice, we train a separate TTS model (five in total).
% These models will be used as a backbone of proposed multilingual TTS system for Turkic languages.
These models will be used as the backbone of the proposed multilingual TTS system for Turkic languages.

% To enable speech synthesis from other Turkic languages, we constructed IPA-based converter module.
To enable speech synthesis for other Turkic languages, we constructed an IPA-based conversion module.
% The IPA-based converter takes letters from alphabets of other Turkic languages and converts them to the letters of Kazakh alphabet\footnote{The IPA-based converter is guaranteed to be one-to-one mapping for Kazakh language only. In other words, the text obtained by converting back from IPA symbols will match the original input text only for Kazakh, while for other languages they might not match.}.
The IPA-based converter takes letters from the alphabets of other Turkic languages and converts them into the letters of the Kazakh alphabet.
%\footnote{Note that the IPA-based converter guarantees a one-to-one mapping for Kazakh only. In other words, the text obtained by converting back from IPA symbols matches the original input text only for Kazakh, while it may not match for other languages.}.
% This is achieved by first converting the input letters into corresponding IPA representations.
For this purpose, the letters entered are first converted into the corresponding IPA representations.
%%% SOUNDS OUT OF CONTEXT
%\textcolor{red}{The IPA is a set of symbols and rules widely used to represent spoken languages at the phonetic level.}
% Next, the IPA symbols are converted to the letters of Kazakh alphabet, which can be used as an input to constructed TTS models.
Next, the IPA symbols are converted into the letters of the Kazakh alphabet, which can be used as input for the TTS models constructed.

%The mappings of the Turkic alphabets onto IPA symbols are given in Table~\ref{tab:ipa}. 
% It was manually created based on our expertise as we couldn't find complete mapping that would enable Turkic to Kazakh conversion without any flaws and covering all addressed languages.
%These were manually created based on our expertise, as we could not find a complete mapping that would allow an error-free conversion from Turkic to Kazakh and cover all the languages addressed.

The mappings of the Turkic alphabets onto IPA symbols were manually created based on our expertise, as we could not find a complete mapping that would allow an error-free conversion from Turkic to Kazakh and cover all the languages addressed.
% Since the Kazakh is used as a source languages, we selected only 42 IPA symbols corresponding to 42 letters of Kazakh alphabet.
Since Kazakh is used as a source language, we selected only 42 IPA symbols corresponding to the 42 letters of the Kazakh alphabet.
It is worth mentioning that, of the Turkic languages in question, Kazakh---along with Bashkir---has the most letters and contains a large majority of the phonemes of the target languages.
% The developed mappings can be also used as a guide for other works aiming to build Turkic speech recognition and speech translation systems. 
The developed mappings can also be used as a guide for other work aimed at building multilingual systems for Turkic languages, such as speech recognition, speech translation, and so on.
% Due to space constraints, we provide the mappings of the Turkic alphabets onto IPA symbols in our GitHub repository\footnote{\href{https://github.com/IS2AI/TurkicTTS}{https://github.com/IS2AI/TurkicTTS}}.
% Due to space constraints, we provide the mappings of the Turkic alphabets onto IPA symbols in our GitHub repository\textsuperscript{\ref{ft:github}}.
% \vspace{-0.2cm}
%%%%% REPLACED %%%%%% Due to space constraints, we provide the mappings of the Turkic alphabets onto IPA symbols in our GitHub repository\textsuperscript{\ref{ft:github}}.
The mapping of the Turkic alphabets onto IPA symbols is provided in Table~\ref{tab:ipa}.
\vspace{-0.2cm}

\iffalse
It is important to emphasize, that in Uyghur language, two witting systems are widely used\footnote{In fact, Cyrillic-derived alphabet for Uyghur is also present, which is still in use in some former Soviet republics.}: Latin and Arabic.
In our work, we favored the Latin as it was easier for us to correlate it's letters with other Turkic languages.
\fi

\iffalse
\subsection{IPA conversion}
\label{sec:ipa}
%What is IPA, why we need them, where do we use it? 1-3 sentences
The IPA is a set of symbols and rules widely used to represent spoken languages at the phonetic level. Originally, the IPA was devised in the 1880s by teachers of English, French, and German as a standardised system for the phonetic transcription of these languages. Today, the IPA is a standard guide to pronunciation and a useful tool for practical linguistic purposes, including, inter alia, transcribing the sounds of the world's oral languages, developing writing systems for previously unwritten languages, and learning foreign languages.

In this work, we address ten Turkic languages bla-bla.
The characteristics of those langauges are given in Table 1. bla-bla.
The IPA mapping used in our work is shown in Table 2 bla-bla.
bla-bla.
For Uyghur we used Latin version which is employed by X\% of Uyghur speakers, mostly in bla-bla region.
bla-bla.
\fi

% \iffalse
\begin{table*}[!ht]\scriptsize
    \begin{center}
        \caption{The letter-to-symbol mapping used in the study}\label{tab:ipa}
        % \vspace{0.1cm}
        \renewcommand{\arraystretch}{0.95}%
        \begin{tabular}{c|ccccccccccc}
            \toprule
            \textbf{IPA} & \textbf{Azerbaijani} & \textbf{Bashkir} & \textbf{Kazakh} & \textbf{Kyrgyz} & \textbf{Sakha} & \textbf{Tatar} & \textbf{Turkish} & \textbf{Turkmen} & \textbf{Uyghur} & \textbf{Uzbek} \\ 
            \midrule
            \textscripta & a & а & а & а & а & а & a & a & a & - \\
            æ & \textschwa & ә & ә & - & - & ә & - & ä & e & a \\
            b & b & б & б & б & б & б & b & b & b & b \\
            v & v & в & в & в & в & в & v & w & w & v \\
            g & q & г & г & г & г & г & g & g & g & g \\
            \textgamma & ğ & ғ & ғ & - & \cyrghk & - & ğ & - & gh & g\textipa{‘} \\
            d & d & д & д & д & д & д & d & d & d & d \\
            \t{je} & - & е & е & е & е & е & - & - & - & - \\
            \t{j\textopeno} & - & ё & ё & ё & ё & ё & - & - & - & - \\
            \textyogh & j & ж & ж & - & ж & ж & j & ž & zh & - \\
            z & z & з & з & з & з & з & z & - & z & z \\
            \t{ij} & i & и & и & и & и & и & i & i & i & i \\
            j & y & й & й & й & й & й & y & ý & y & y \\
            k & k & к & к & к & к & к & k & k & k & k \\
            q & - & ҡ & қ & - & - & - & - & - & q & q \\
            l & l & л & л & л & л & л & l & l & l & l \\
            m & m & м & м & м & м & м & m & m & m & m \\
            n & n & н & н & н & н & н & n & n & n & n \\
            \textipa{\ng} & - & ң & ң & ң & ҥ & ң & - & ň & ng & ng \\
            \textopeno & o & о & о & о & о & о & o & o & o & o \\
            \textbaro & ö & ө & ө & ө & ө & ө & ö & ö & ö & o\textipa{‘} \\
            p & p & п & п & п & п & п & p & p & p & p \\
            r & r & р & р & р & р & р & r & r & r & r \\
            s & s & с & с & с & с & с & s & - & s & s \\
            t & t & т & т & т & т & т & t & t & t & t \\
            \t{\textupsilon\textipa{w}} & u & у & у & у & у & у & u & u & u & u \\
            \textupsilon & - & - & ұ & - & - & - & - & - & - & - \\
            \textscy & ü & ү & ү & ү & ү & ү & ü & ü & ü & - \\
            f & f & ф & ф & ф & ф & ф & f & f & f & f \\
            x & х & х & х & х & х & х & - & - & x & x \\
            h & h & һ & һ & - & һ & һ & h & h & h & h \\
            \t{ts} & - & ц & ц & ц & ц & ц & - & - & - & - \\
            \t{t\textesh} & ç & ч & ч & ч & ч & ч & ç & ç & ch & ch \\
            \textesh & ş & ш & ш & ш & ш & ш & ş & ş & sh & sh \\
            \textctc & - & щ & щ & щ & щ & щ & - & - & - & - \\
            \textglotstop & - & ъ & ъ & ъ & ъ & ъ & - & - & - & \textipa{'} \\
            \textramshorns & ı & ы & ы & ы & ы & ы & ı & y & - & - \\
            \textsci & - & - & і & - & - & - & - & - & - & - \\
            \textsuperscript j & - & ь & ь & ь & ь & ь & - & - & - & - \\
            e & e & э & э & э & э & э & e & e & ё & e \\
            \t{j\t{\textupsilon\textipa{w}}} & - & ю & ю & ю & ю & ю & - & - & - & - \\
            \t{j\textscripta} & - & я & я & я & я & я & - & - & - & - \\
            \t{d\textyogh} & c & - & - & ж & дь & җ & c & j & j & j \\
            g\textsuperscript{j} & g & - & - & - & - & - & - & - & - & - \\
            \texttheta & - & ҫ & - & - & - & - & - & s & - & - \\
            \textipa{\dh} & - & ҙ & - & - & - & - & - & z & - & - \\
            \textltailn & - & - & - & - & нь & - & - & - & - & - \\
            \midrule
            \textbf{47} & \textbf{32} & \textbf{42} & \textbf{42} & \textbf{36} & \textbf{40} & \textbf{39} & \textbf{29} & \textbf{30} & \textbf{32} & \textbf{30} \\
            \bottomrule
        \end{tabular}
    \end{center}
    \vspace{-0.5cm}
\end{table*}
% \fi

\section{Experimental setup}
% \vspace{-0.2cm}
\label{sec:exp_setup}
\iffalse
We can try two approaches:

1) text -> Turkic-to-IPA -> IPA-based TTS -> synthesized sound

2) text -> Turkic-to-IPA -> IPA-to-Kazakh -> Kazakh TTS -> synthesized sound
\fi

\subsection{Source language data}
% In our experiments, Kazakh was used as a source languages, and we employed KazakhTTS2 corpus.
In our experiments, Kazakh was used as a source language, and we employed the KazakhTTS2 speech corpus.
% As was mentioned earlier, we trained five TTS models corresponding to five voices present in KazakhTTS2.
As mentioned earlier, we trained five TTS models corresponding to the five voices present in the corpus.
% All five models were verified to be cable of synthesizing Turkic speech.
% However, we used only \textbf{Speaker X}'s voice to evaluate the proposed approach due to the time constraints. 
All five models proved capable of synthesising Turkic speech.
% However, we used only \textbf{Speaker X}'s voice to evaluate the proposed approach due to the time constraints. 
%However, due to the time constraints, we only used the voice of Speaker M2 to evaluate the proposed approach.
However, based on our previous experience, we found that raters are usually unwilling to participate in or complete an overly long evaluation session.
Therefore, we only used the voice of Speaker M2 to evaluate the proposed approach.
% Moreover, based on our previous experience, we observed that raters are usually unwilling to participate or fully complete the evaluation session if it is too long.
%Moreover, based on our previous experience, we found that raters are usually unwilling to participate in or complete an overly long evaluation session.
% The \textbf{Speaker X}'s data covers around \textbf{42.5} thousand unique tokens and over \textbf{24 thousand} utterances with a mean segment duration of \textbf{8.3} seconds.
Speaker M2's data contain approximately 58 hours of transcribed speech. % consisting of over 32 thousand utterances.
%The mean utterance duration is around 6.3 seconds, and the number of unique tokens is around 53 thousand.
%\textbf{Speaker X}'s data include around \textbf{53} thousand unique tokens and over \textbf{32 thousand} utterances with a mean duration of \textbf{6.3} seconds.
% We did not use any data for the target languages to simulate zero-shot learning scenario.
We did not use any data for the target languages to comply with the zero-shot learning scenario.

\iffalse
% Specifically, we used Speaker M1's News data consisting of 57 hours of transcribed data.
Specifically, we used \textbf{Speaker M1}'s news data consisting of 57 hours of transcribed speech read by a male.
% This portion of corpus covers around 42.5 thousand unique tokens and over 24 thousand segments with the mean segment duration equal to 8.3 seconds. 
This portion of the corpus covers around 42.5 thousand unique tokens and over 24 thousand segments with a mean segment duration of 8.3 seconds. 
% For the target languages, we did not use any linguistic resources.
We actually validated that all five voices are suitable for synthesizing Turkic speech, however, we decided to evaluate only one voice due to the time constraints.
Moreover, based on our previous experience, we observed that raters are usually unwilling to participate in a long evaluations.
We did not use any data for the target languages to simulate zero-shot lerarning scenario.
\fi

% \vspace{-0.2cm}
\subsection{TTS architecture}
% In our experiments, we employed end-to-end TTS system based on the Tacotron~2 \cite{DBLP:conf/icassp/ShenPWSJYCZWRSA18} architecture.
In our experiments, we trained an E2E TTS model based on the Tacotron~2~\cite{DBLP:conf/icassp/ShenPWSJYCZWRSA18} architecture using the NVIDIA DGX A100 machines.
% Specifically, we followed LJ Speech~\cite{ljspeech17} training recipe implemented within ESPnet-TTS toolkit~\cite{hayashi2020espnet}.
Specifically, we followed the LJ Speech~\cite{ljspeech17} training recipe implemented within the ESPnet-TTS toolkit~\cite{hayashi2020espnet}.
% All TTS models were trained using NVIDIA DGX A100 machines.
% We trained two TTS models: IPA-based TTS and Kazakh TTS.'
In~\cite{mussakhojayeva2022kazakhtts2}, the speakers were asked to pay attention to punctuation by pausing at commas and using the right intonation when pronouncing sentences ending with question marks and exclamation points. In order to use the intonation and pacing considered in~\cite{mussakhojayeva2022kazakhtts2}, we felt that the input for our TTS model should also include five punctuation symbols (‘.’, ‘,’, ‘-’,‘?’, ‘!’) in addition to a text sequence of 42 Kazakh letters.
% The input for the TTS model was a text sequence consisting of 42 Kazakh letters and five punctuation symbols (‘.’, ‘,’, ‘-’,‘?’, ‘!’).
The output was a sequence of acoustic features (80 dimensional log Mel-filter bank features).
To transform these acoustic features into time-domain waveform samples, we employed WaveGAN~\cite{DBLP:conf/icassp/YamamotoSK20} vocoder. 

\iffalse
We trained two TTS models—an IPA-based TTS and a Kazakh TTS—using NVIDIA DGX A100 machines.
% The input for IPA-based TTS model is a sequence of IPA symbols (42 symbols) and 5 punctuation symbols (‘.’, ‘,’, ‘-’,‘?’, ‘!’).
The input for the IPA-based TTS model was a sequence of 42 IPA symbols and five punctuation symbols (‘.’, ‘,’, ‘-’,‘?’, ‘!’).
% The input for Kazakh TTS model 5 punctuation symbols, same as in IPA-based TTS, and a sequence of Kazakh letters (42 letters).
% The input for the Kazakh TTS model was five punctuation symbols, same as in IPA-based TTS, and a sequence of Kazakh letters (42 letters).
Similar to the IPA-based TTS, the input for the Kazakh TTS model was five punctuation symbols and a sequence of 42 Kazakh letters.
% For both models, the output is a sequence of acoustic features (80 dimensional log Mel-filter bank features).
For both models, the output was a sequence of acoustic features (80 dimensional log Mel-filter bank features).
% To transform these acoustic features into the time-domain waveform samples, we employed WaveGAN~\cite{DBLP:conf/icassp/YamamotoSK20} vocoders. 
To transform these acoustic features into time-domain waveform samples, we employed WaveGAN~\cite{DBLP:conf/icassp/YamamotoSK20} vocoders. 
\fi

% In the Tacotron 2 model, the encoder module was modeled as a single bidirectional LSTM layer with 512 units (256 units in each direction), and the decoder module was modeled as a stack of two unidirectional LSTM layers with 1,024 units.
In the Tacotron 2-based TTS system, the encoder module was modelled as a single bidirectional LSTM layer with 512 units (256 units in each direction), and the decoder module was modelled as a stack of two unidirectional LSTM layers with 1,024 units.
The parameters were optimised using the Adam algorithm with an initial learning rate of $10^{-3}$ for 200 epochs. 
To mitigate overfitting, we applied a dropout of 0.5.
%A separate Tacotron 2 model was trained for each speaker (i.e., a single speaker model).
% More details on the model specifications and training procedures can be found in our GitHub repository\textsuperscript{\ref{ft:github}}.
More details on the model specifications and training procedures can be found in our GitHub repository\textsuperscript{\ref{ft:github}}.

\subsection{Evaluation process}
% To assess the quality of the synthesised recordings, we conducted subjective evaluation using the MOS measure. 
% To assess the quality of the synthesised recordings, we performed a subjective evaluation using the Qualtrics survey platform~\cite{qualtrics}.
To assess the quality of the synthesised recordings, we conducted a subjective evaluation using an online survey on the Qualtrics XM Platform\footnote{\url{https://qualtrics.com}}.
% The platform enables data to be collected automatically through a number of web services and aggregates participants' responses in user-friendly databases, tables and charts.
% To recruit volunteer raters, we promoted our work on popular YouTube channels operating in the Turkic languages.
%To recruit volunteer raters, we distributed the link to the survey on popular YouTube channels and social media pages operating in the Turkic languages.
To recruit volunteer raters, we distributed the link to the survey on popular social media platforms operating in the Turkic languages.
% We prepared a separate evaluation session for each addressed language.
% For each target language, we prepared a separate evaluation survey.
We created a separate evaluation questionnaire for each target language.

% The evaluation session first presented welcome message including general information introducing the research authors, aim, etc.
%%% The evaluation survey began with a welcome message that set out the context in which the project was undertaken and outlined the details of what raters were expected to do in the survey and how long completing it might take.%%%
% Next, raters were asked to read and agree with the rules and responsibilities listed in a consent form, and fill in information about their age, gender, and bla-bla.
% Informed consent was obtained from raters, certifying they were 18 years old or above and confirming their participation in the project.
Informed consent was obtained from raters, certifying they were at least 18 years old and confirming their participation in the project.
% The raters aged below 18 were not allowed to participate.
Individuals younger than 18 years old were not allowed to participate.
%Raters were also invited to fill in a questionnaire about their age, gender, and native language.
The survey was first developed in English and later translated into the Turkic languages with the help of Qualtrics and other online translation services.
%The session was developed in L1, L2, etc. 
%First, we designed an English version and later translated it into L1, L2 etc using multilingual neural machine translation services (Google, Yandex etc)/Qualtrics.
%%The survey for each language was inspected by native speaker.%%
%However, for the sake of confidence, we back-translated the survey content into English, and verified its correspondence with the original English version.
For some languages, we also provided the instructions in English and Russian.
%For the sake of confidence, we back-translated the information sheet, consent form, task descriptions etc into English. 
%The original English and the back-translated versions appeared almost identical, for which reason the versions were deemed adequate to proceed with distribution.

% Overall, the evaluation session consisted of three main parts, each assessing different aspects of synthesized speech.
The evaluation survey consisted of three main parts, each assessing different aspects of the synthesised speech.
% The first part assessed the overall quality of synthesized speech, where ten recordings with the accompanying transcript were presented to raters.
The first part assessed the overall quality of the synthesised speech, with raters presented with ten recordings and their transcripts.
% The raters were instructed to play the recordings using headphones in a quiet environment, and rate them using a five-point Likert scale: 5 for excellent, 4 for good, 3 for fair, 2 for poor, and 1 for bad.
%% Raters were instructed to listen to the recordings through headphones in a quiet environment and rate them on a five-point Likert scale: 5 for excellent, 4 for good, 3 for fair, 2 for poor, and 1 for bad.%%
Raters were instructed to listen to the recordings and rate them on a five-point Likert scale~\cite{likert1932technique}: 5 for \textit{excellent}, 4 for \textit{good}, 3 for \textit{fair}, 2 for \textit{poor}, and 1 for \textit{bad}.
% They were allowed to listen to the recordings several times but were not allowed to alter the ratings once submitted.
They were allowed to listen to the recordings several times, but could not alter the ratings once submitted.
% The evaluation recordings were presented one at a time and in the same order for all raters.
The evaluation recordings were presented to all raters one at a time and in the same order.

% The second part evaluated the comprehensibility aspect of the synthesized speech.
The second part assessed the comprehensibility aspect of the synthesised speech.
% bla-bla.
To reduce cognitive effort, raters were presented with five straightforward multiple-choice questions. There were four options per question, with only one being the correct answer.
% The questions were between \textbf{X-Y} words long; the options consisted of one word.
The questions were on average 4.84 words long; the options consisted of one or two short words.
% Each question constituted one score, and we will report the average score for each langauge (e.g., 3.5 out of 5)
A score of one was attached to the right answer.

% Lastly, the third part evaluated the intelligibility aspect.
Finally, the third part assessed the intelligibility aspect of the synthesised speech by using semantically unpredictable sentences (SUS)~\cite{benoit1996sus}.
% The SUS were constructed using 23-25 commonly used words.
The SUS were formed from 23-25 commonly used words.
% Raters listen to a recording, constructed using SUS approach and consisting of 5-6 words, and attempt to transcribe it.
Raters were asked to listen to five sentences and write down what they heard in a field.
Raters were informed that the sentences were not meaningful although they contained real words.
% The intelligibility was evaluated using word error rate (WER) measure, where we compared the original text with the one submitted by raters.
To evaluate intelligibility, only the sentences that were entirely correct were considered.
% We will report CER and WER for each language.
%% After completion of the third task, participants were thanked for participation and asked to end the survey.%%
\vspace{-0.2cm}

\section{Evaluation results}
\label{sec:results}
% \vspace{-0.2cm}
The survey results are given in Table~\ref{tab:results}.
% In total, for each language, at least X raters participated with the \textbf{Kazakh} languages having the maximum number of raters (X) followed by \textbf{Turkish} (Y) and \textbf{Uzbek} (Z).
Of 572 registered survey participants, 24 were under the age of 18. Raters varied in terms of both gender and age. The number of raters who answered all survey questions was 435, with at least six participating in a survey per language. 
The highest number of raters, 254, was observed for Sakha, followed by 151 and 47 raters for Kazakh and Azerbaijani, respectively.
% The questionnaire results showed that the raters varied in gender, but not in age (most of them were under 20).
% The questionnaire results showed that raters varied in gender, but not in age (most of them were under 20).
%Specifically, the majority of raters were from the south and west of Kazakhstan, and females outnumbered males by a factor of 1.5.
% In addition, some information about participants bla-bla.

% Evaluation results for three parts are given in Table~\ref{tab:results}.

% The MOS on overall quality is within 2.77-4.18 range, with the Kazakh language achieving the highest score as expected.
The MOS for overall quality ranges from 2.37 to 4.18, with Kazakh scoring the highest, as expected.
% The best MOS among target languages are achieved by Kyrgyz, Turkish, and Uyghur languages.
The best scores among the target languages were achieved by Kyrgyz, Turkish, and Uyghur. This is remarkable, given the quality of the synthesised recordings was evaluated as above \textit{fair} by speakers of the languages belonging to three Turkic branches.
%We presume that the reason is because they belong to the same family sub-branch as Kazakh bla-bla.
% The worst scores are achieved for Azerbaijani and Bashkir languages.
% The worst results were obtained by Azerbaijani and Bashkir, with the quality of the recordings rated as just below average.
%The worst performers were Azerbaijani and Bashkir, where the quality of the recordings was rated just below average.
The worst performer was Turkmen, where the quality of the recordings was rated just below average.

% In comprehensibility test, the highest score among target languages is achieved in X language, where raters on average correctly answered X questions.
% In the comprehensibility test, Turkish scored the highest among the target languages, as the raters answered 96\% of the questions correctly.
In the comprehensibility test, Sakha, Bashkir, Turkish, and Azerbaijani scored the highest among the target languages, with at least 90\% of the answers being right.
% While the worst score is achieved in languages X, Y, and Z.
The lowest score was obtained by Uyghur, with raters answering only every second question correctly.
% Comparing the results of the two tasks, it can be seen that although Sakha speakers---constituting the majority of the survey participants---evaluated the quality of the synthesised recordings at a MOS of only 2.85, their responses in the comprehensibility test were 93\% right.
Comparing the results of the two tasks, it can be seen that although Sakha speakers---constituting the majority of the survey respondents---evaluated the quality of the synthesised recordings at a MOS of only 2.85, their responses in the comprehensibility test were 93\% right.
This seems to indicate that recordings synthesised using a Kazakh voice can be relatively easy to understand for speakers of the Siberian Turkic languages.
%It is important to mention that five out of ten languages achieved 90\% or above in this test.

% In intelligibility test, the highest score among target languages is achieved in X and Y languages, and the worst score is achieved in Z language.
In the intelligibility test, Turkish (61\%) and Turkmen (57\%) scored highest among the target languages, while Sakha obtained a result of only 15\%. As expected, writing down SUS was the most challenging task, probably due to the greater cognitive load required to correctly recognise a complete sequence of semantically unrelated words. This is consistent with~\cite{benoit1996sus}, where intelligibility scores were in the range of 10-20\%.
% \textcolor{red}{The evaluation results in three parts are highly correlated, which shows bla-bla.}

Overall, the obtained results appear promising and sufficient for most real-world applications.
Moreover, we believe that performance can be further improved for all target languages by fine-tuning the pre-trained models with the seed data of the corresponding language.
%\vspace{-0.2cm}

\begin{table}[t]
    \begin{center}
        % \small
        %\renewcommand\arraystretch{1.1}
        %\setlength{\tabcolsep}{1.0mm}
        % \vspace{-0.225cm}
        \caption{The survey statistics for rater number (R), gender (F \& M), and age ( < 45 \& 45+) and the evaluation results of the overall quality (Q), comprehensibility (C), and intelligibility (I) of synthesised speech}\label{tab:results}
        % \vspace{-0.25cm}
        \resizebox{\columnwidth}{!}{
        \begin{tabular}{l|ccccc|ccc}
            \toprule
            \textbf{Language}   &
            \textbf{R}  &
            \textbf{F}  &
            \textbf{M}  &
            \textbf{< 45}  &
            \textbf{45+}  &
            \textbf{Q}  & \textbf{C} & \textbf{I}\\
            \midrule
            Azerbaijani & 47 & 22 & 25 & 22 & 25        & 2.93 & 90\%                      & 52\% \\ 
            Bashkir & 11 & 8 & 3 & 4 & 7             & 2.67              & 92\%                      & 47\% \\ 
            Kazakh & 151 & 89 & 62 & 120 & 31              & 4.18              & 97\%                      & 80\% \\ 
            Kyrgyz & 14 & 12 & 2 & 6 & 8              & 3.54              & 86\%                      & 43\% \\ 
            Sakha & 254 & 155 & 99 & 147 & 107               & 2.85              & 93\%                      & 15\% \\ 
            Tatar & 15 & 12 & 3 & 3 & 12               & 2.82              & 79\%                      & 17\% \\ 
            Turkish & 18 & 6 & 12 & 15 & 3             & 3.25              & 91\%                      & 61\% \\ 
            Turkmen & 6 & 0 & 6 & 6 & 0             & 2.37              & 67\%                      & 57\% \\ 
            Uyghur & 10 & 6 & 4 & 6 & 4              & 3.01              & 45\%                      & 26\% \\ 
            Uzbek & 22 & 2 & 20 & 19 & 3               & 2.85              & 80\%                      & 45\% \\
            \midrule
            \textbf{Total} & 548 & 312 & 236 & 348 & 200 & 3.25 & 92\% & 41\%\\
            \bottomrule
        \end{tabular}
        }
        %\vspace{0.1cm}
        %{\raggedright \textit{Note.} The news source data of speakers F1 and M1 are from KazakhTTS.\par}
    \end{center}
\vspace{-0.6cm}
\end{table}

% \vspace{-0.3cm}
\section{Challenges and Future Work}
\label{sec:discuss}
% \vspace{-0.2cm}
% During this work, we faced several challenges that the future work should be aware of.
In this study, we faced several challenges that should be considered in future work.
The first and probably the most important one is the scarcity of speech corpora for TTS in Turkic languages.
% We observed that most of the existing datasets are proprietary, whereas majority of available datasets are either of low-quality or unsuitable for building the state-of-the-art TTS architectures.
We found that most of the existing datasets are proprietary, while the majority of available datasets are either of low quality or unsuitable for building state-of-the-art TTS architectures.
To facilitate the research and development of TTS systems for Turkic languages, future studies should focus on collecting high-quality and open-source speech corpora.
Furthermore, Turkic languages are agglutinative, with rich vocabularies and many characters per word, which requires 
that the datasets collected be large in size.
% Therefore, collected datasets should be of large-scale.

Another challenge is loanwords and code-switching.
% Specifically, the Turkic languages used in countries which were part of Soviet union are frequently mixed with the Russian and have a lot of words borrowed from it.
Specifically, the Turkic languages spoken in the former-Soviet countries are often used along with Russian and thus contain many Russian borrowings.
Usually, these words retain the orthographic and phonological properties of the original language.
Consequently, this might mislead TTS systems, as Russian is different from Turkic languages in many aspects~\cite{mussakhojayeva2022kazakhtts2}.
%For example, in Russian, the stress can be on any syllable of a word, whereas most words in Turkic languages are stressed in final syllable.
%Furthermore, the spelling of Turkic words usually closely matches their pronunciation, which is not the case with Russian words; for example, the letter ``\textit{o}'' is sometimes pronounced as \textipa{/a/}.
%In general, the number of loanwords from other languages, especially English, is also increasing, which is likely to pose an additional challenge in the near future.

%\textbf{Homoglyph issue.} Common among Turkic languages that shifted from Cyrillic to Latin.
% We also encountered homoglyph issue almost in all Turkic languages that shifted from Cyrillic to Latin alphabet, where visually similar characters from two scripts are used interchangeably.
We also observed the interchangeable use of visually similar/identical characters from different scripts---also known as homoglyphs---in almost all Turkic languages that have transitioned from the Cyrillic to the Latin alphabet.
%\textbf{For example, give one or two simple examples bla-bla.}
% Usually this will not disturb a reader, but might cause a serious problem to the computer systems. %, such as TTS, ASR, and speech translation.
Normally, this does not bother the reader, but can be problematic for computer systems. 
% Therefore, collected text should be carefully inspected.
Therefore, the collected text should be carefully inspected.

% Finally, we did not measure the literacy level of raters prior to the intelligibility test. Neither, did we take into account that raters may not have the keyboard layout of the native language on their devices. 
Finally, we did not measure the literacy skills of raters; nor did we ensure that raters had the native language keyboard layout on their devices. Information about raters' literacy levels and the provision of a virtual keyboard layout of the required language may have helped us establish a greater degree of accuracy in the intelligibility test.

% We believe that overcoming these challenges for the Turkic language will be an interesting direction for future research.
We believe that addressing these challenges for the Turkic languages will be an interesting direction for future research.
% We hope that our work will encourage subsequent efforts in this area to address some of the practical issues that arise when training TTS systems for the Turkic language.
We hope that our work will encourage subsequent efforts in this area to solve some of the practical problems encountered in training TTS systems for the Turkic languages.
% We also hope that our proposed approach will serve as a baseline for future research and as a preliminary solution for real-world applications.
We also hope that our proposed approach will serve as a baseline and as a preliminary solution for real-world applications.
% Besides, the future work should expand to other Turkic languages not considered in this paper.
We will consider comparative experiments, as we are interested in a more detailed evaluation of the performance of the model, and explore the advantages of using Kazakh speech to build TTS models for Turkic languages.
% Besides, future work should be extended to other Turkic languages that were not considered in this work.
In addition, future work should be extended to other Turkic languages not considered in this work.
%\vspace{-0.2cm}

% \vspace{-0.3cm}
\section{Conclusion}
% \vspace{-0.2cm}
\label{sec:conc}
In this work, we developed a multilingual TTS system for ten Turkic languages.
We assumed a zero-shot learning scenario where no target data were used.
The proposed approach employed a TTS system trained using Kazakh and an IPA-based converter to translate letters from the target languages into the source language.
% When evaluating the quality of the synthesised speech, a MOS of 3.54 was achieved for Kyrgyz.
When evaluating the quality of the synthesised speech over all addressed languges, a MOS of 3.25 was achieved.
% In the comprehensibility test, 96\% of the Turkish raters gave correct answers.
In the comprehensibility test, 92\% of the answers were correct.
% In the intelligibility test, 67\% of the Bashkir sentences matched the references.
In the intelligibility test, 41\% of the sentences matched the references.
%% To the best of our knowledge, our work is the first attempt to build an E2E TTS system for most of the target languages.
% We evaluated the developed multilingual TTS system using three measures: overall quality, comprehensibility, and intelligibility.
%% We evaluated the system using three measures: overall quality, comprehensibility, and intelligibility.
%% Considering the linguistic similarity of Turkic languages, the results obtained are promising and should be sufficient for most real-world applications. %, especially for X, Y, and Z languages.
% Moreover, our approach can be used as a strong baseline for future work, and further adapted to the target language using some seed data.
%% Moreover, our approach can serve as a strong baseline for future work and can be further adapted to the target language using some seed data.
Given that this is the first attempt at building a Turkic TTS system, the achieved results are promising.
% To enable experiment reproducibility, we share our code, pre-trained models, and dataset in our GitHub repository\textsuperscript{\ref{ft:github}}.
To enable experiment reproducibility, we share our code, pre-trained models, and dataset in our GitHub repository\textsuperscript{\ref{ft:github}}.
%This is the first attempt to build multilingual Turkic TTS system, and we strongly believe that our work will help to advance speech processing research for understudied Turkic languages.
%\vspace{-0.2cm}

\section{Acknowledgements}
We would like to extend our heartfelt thanks to all individuals who contributed to the recruitment of participants for this study. Their efforts were critical to the success of our survey. In particular, we would like to express our deepest appreciation to Viktor Krivogornitsyn for his extraordinary dedication in attracting a substantial number of Sakha speakers. His contribution was invaluable, and we are grateful for his support.

\bibliographystyle{IEEEtran}
\bibliography{main}

\end{document}